\documentclass[preprint,prb,showpacs,superscriptaddress]{revtex4}

\usepackage{graphicx}% Include figure files
\usepackage{dcolumn}% Align table columns on decimal point
\usepackage{bm}% bold math
\usepackage{amsmath}

\begin{document}

\title{Tuning polarization and elasticity properties by uniaxial stress in BiFeO$_{3}$}

\author{Huafeng Dong}
\affiliation{State Key Laboratory of Superlattices and Microstructures, Institute of Semiconductors, Chinese Academy of Sciences, P.O. Box 912,Beijing 100083, China}
\affiliation{Department of Physics, Tsinghua University, Beijing 100084, China}
\author{Jingbo Li}
\email{jbli@semi.ac.cn}
\affiliation{State Key Laboratory of Superlattices and Microstructures, Institute of Semiconductors, Chinese Academy of Sciences, P.O. Box 912,Beijing 100083, China}
\author{Zhigang Wu}
\email{zhiwu@mines.edu}
\affiliation{Department of Physics, Colorado School of Mines, 1523 Illinois Street, Golden, CO 80401, USA}

\date{\today}

\begin{abstract}
The changes of polarizational, elastic and structural properties of mutiferroic BiFeO$_{3}$ under [111] direction uniaxial stress are calculated using density functional theory within the Perdew-Burke-Ernzerhof revised for solids (PBEsol) + $U$ approach, and compared with available measurements or predictions. A large ferroelectric polarization of 87.5 $\mu \texttt{C}/\texttt{cm}^{2}$ is found, agreeing with other theoretical and experimental values, and the polarization changes near-linearly within the uniaxial stress range of 8 GPa to -8 GPa. This property can be used to indirectly control the radiative recombination of luminous organ and the electrical properties of transistor structure. In addition, we have also investigated the elastic stiffness coefficients and the elastic compliance coefficients in the same uniaxial stress range, which provide helpful guidance for the applications of BiFeO$_{3}$.

\end{abstract}

\pacs{77.84.-s, 77.22.Ej, 62.20.D-}

\maketitle

\section{Introduction}

Multiferroics are promising materials because of their fundamental physical properties as well as their wide range of potential applications \cite{r1,r2,r3}. Among these materials, BiFeO$_{3}$ is the most promising multiferroics from the applications perspective \cite{r3}, and the latest results, such as photovoltaic effect \cite{r4} and ferroelectric-resistive random access memory \cite{r5}, it offers continue to fuel the interest in it.

At room temperature, bulk BiFeO$_{3}$ has long been known to be ferroelectric, with a Curie temperature $\sim$1110 K, and antiferromagnetic, with a N\'{e}el temperature $\sim$643 K \cite{r6,r7,r8}. It is a rhombohedral $R3c$ perovskite with a large electric polarization \cite{r9} 88.7 $\mu \texttt{C}/\texttt{cm}^{2}$ pointing along the [111] direction, and Bi, Fe, and O are displaced relative to one another along this threefold axis. The largest relative displacements are those of Bi relative to O, consistent with a stereochemically active Bi lone pair which induces a structural distortion that can lead to ferroelectricity \cite{r9}. But under high pressure (i.e. strain), the Bi atom tend to form planar BiO$_{3}$ groups and the Fe$^{3+}$ show a high spin (HS) - low spin (LS) crossover transition at pressures around 40 GPa \cite{r10}. Furthermore, when the rhombohedral perovskite BiFeO$_{3}$ is grown on a substrate having a square in-plane lattice, its symmetry is lowered to a monoclinic phase \cite{r11}. As the epitaxial strain increaseing, it undergoes a phase sequence of rhombohedral (\emph{R}) to monoclinic (\emph{R}-like \emph{M}$_{A}$) to monoclinic (\emph{T}-like \emph{M}$_{C}$) to tetragonal (\emph{T}), which is otherwise seen only near morphotropic phase boundaries (MPB) in lead-based solid-solution perovskites \cite{r12}. The above results thus show that ``strain engineering'' is an important tool both in fundamental studies to understand complex phase transition as well as in approaches to find new, lead-free materials with technologically relevant properties, such as large piezoelectric coefficients and tunable polarization. In this work, we report the changes of polarization and elasticity properties under uniaxial stress (strain) in BiFeO$_{3}$.

Elastic strain is an effective way to control the magnetic and the polarization of BiFeO$_{3}$ \cite{r6,r7,r8,r10,r11,r12}. Ramazanoglu {\it et al.} \cite{r6} showed that small uniaxial pressure (i.e. elastic strain) will produce significant changes in the populations of magnetic domains and rotate the magnetic easy plane of domains in BiFeO$_{3}$. Jang {\it et al.} \cite{r13} provided direct experimental evidence that (001)$_{p}$-oriented BiFeO$_{3}$ films exhibit a strong strain tunability of their out-of-plane remanent polarization. However, no theoretical investigation has been done until now to study the changes of polarization and elasticity properties under uniaxial stress in BiFeO$_{3}$, which inhibits the applications of BiFeO$_{3}$ not only in bulk materials but also in thin films. This is another motivation for pursuing this work.

In this work we analyze the changes of polarization, elasticity and structure properties of BiFeO$_{3}$ under [111] direction uniaxial stress and discuss its physical mechanism. It is of great interest to note that the polarization of BiFeO$_{3}$ changes near-linearly in the stress range of 8 GPa to -8 GPa. This property is conducive to the application of BiFeO$_{3}$, such as by controlling the spontaneous polarization to indirectly control the radiant recombination of luminous organ and the electrical properties of transistor structure \cite{r1,r3,r14}. Furthermore, the investigation of elastic and structural properties reveals that the structure of BiFeO$_{3}$ is stable within the uniaxial stress range of 8 GPa to -8 GPa, which provides direction for future applications.

\section{Methodology}

\begin{figure}
%\begin{figure}[tbp]
\begin{center}
\includegraphics[width=\columnwidth]{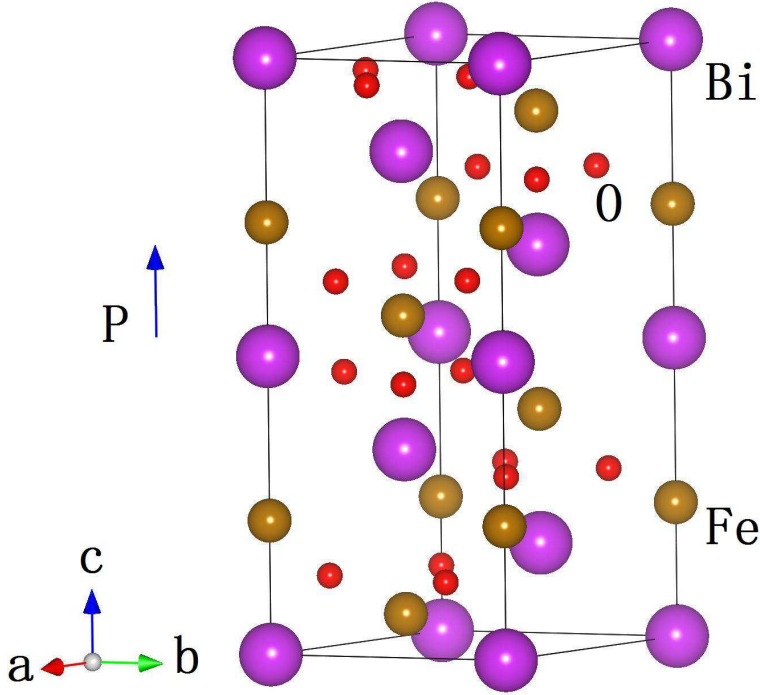}
\end{center}

\caption{\label{fig1}
(Color online) Structure of BiFeO$_{3}$ with hexagonal cell.}
\end{figure}

At room temperature, BiFeO$_{3}$ is a rhombohedral (space group $R3c$) \emph{G}-type antiferromagnetic (AFM) perovskite structure \cite{r3,r15,r16}. In order to exert stress on [111] direction expediently, we do not study the rhombohedral cell ($\alpha = \beta = \gamma \neq 90^{\circ}$) directly but its equivalent one, i.e., the hexagonal cell ($\alpha = \beta = 90^{\circ}, \gamma = 120^{\circ}$) (see Fig. \ref{fig1}) \cite{r17}. The [001] direction of hexagonal cell is the [111] direction of rhombohedral cell. For the sake of calculating the uniaxial stress along the [001] direction, we apply a strain in the [001] direction, and then keep the lattice constant of [001] direction unchanged and relax its perpendicular lattice vectors and all the internal atomic positions. The optimization is done until the other two components of the stress tensor (i.e. $\sigma_{11}$ and $\sigma_{22}$) are all smaller than 0.05 GPa. As a result, with the strain changing continuously we get the variational uniaxial stress.

To calculate the structure, polarization, and elastic coefficient of BiFeO$_{3}$, we use density functional theory within the generalized gradient approximation (GGA) \cite{r18} and GGA+$U$ method \cite{r19} (with the Perdew-Burke-Ernzerhof revised for solids (PBEsol) \cite{r20,r21}) as implemented in the VASP package \cite{r22}. The PBEsol+$U$ method was developed by Dudarev {\it et al.} \cite{r19}, which depended on the effective Coulomb interaction $U_{eff} = U - J$. In the present work, $U_{eff} = 4$ eV is used, as it leads to qualitatively and semiquantitatively correct results for all the properties investigated \cite{r16,r19}. All results are obtained using the projector augmented wave method to represent the ionic cores \cite{r23}, solving for the following electrons: Fe's 3p$^{6}$ 3d$^{6}$ 4s$^{2}$; Bi's 5d$^{10}$ 6s$^{2}$ 6p$^{3}$; and O's 2s$^{2}$ 2p$^{4}$. In the first-principles calculation, we employ a plane-wave cutoff of 500 eV and k-point sampling of 4$\times$4$\times$2 (comparing with k-point sampling of 4$\times$4$\times$8, the error of polarization is less than 0.1\%). The electronic energy is convergent to at least 0.003 meV/atom, and all force components are relaxed to at least 2 meV/{\AA}. The optimizations of cell shape and ion's position are performed by Gaussian smearing technique \cite{r24}. The polarization and the elastic coefficients are calculated by Berry phase \cite{r25,r26} and the finite difference method \cite{r27} respectively.

\section{Results and Discussions}

\subsection{Properties of unstrained BiFeO$_{3}$}

First of all, we study the structure, polarization and the elastic stiffness coefficients of unstrained BiFeO$_{3}$, which are all listed in Table \ref{tab1}. An earlier report \cite{r8} reveals that the ferroelectric properties of BiFeO$_{3}$ are very sensitive to the small changes in the lattice parameters. Consequently, for predicting the ferroelectric properties correctly it is needful to estimate the structural parameters of BiFeO$_{3}$ accurately. As can be seen in Table  \ref{tab1}, our optimized lattice parameters are in good agreement with the experimental values \cite{r28} and other theoretical values  \cite{r29}, which guarantees the veracity of predicted ferroelectric properties and elastic properties. The polarization (87.5 $\mu \texttt{C}/\texttt{cm}^{2}$) is a little bit smaller than the experimental values \cite{r8,r30} but almost equal to the theoretical values (88.7 $\mu \texttt{C}/\texttt{cm}^{2}$) in Ref. [\cite{r9}]. The elastic stiffness coefficients, except for $c_{44}$, are all slightly larger than the values in Ref. [\cite{r7}]. Taking one with another, all the above properties of unstrained BiFeO$_{3}$ are consistent with recent studies \cite{r7,r8,r9,r28,r29,r30,r31}.

\subsection{Near-linearly variational polarization}

\begin{figure}
%\begin{figure}[tbp]
\begin{center}
\includegraphics[width=0.9\columnwidth]{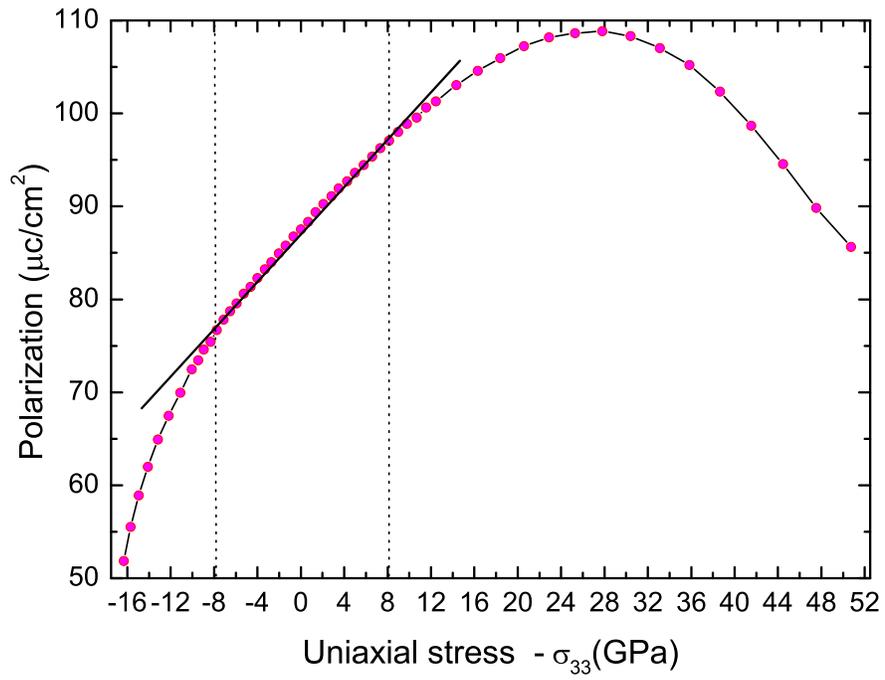}
\end{center}

\caption{\label{fig2}
(Color online) The polarization as a function of uniaxial stress -$\sigma_{33}$.}
\end{figure}

The changes of polarization during the stress changing from 16 GPa (tensile stress) to -52 GPa (compressive stress) are showed in Fig. \ref{fig2}. Interestingly we observed that the polarization $\textbf{P}$ increases almost linearly from 75.45 $\mu \texttt{C}/\texttt{cm}^{2}$ to 97.08 $\mu \texttt{C}/\texttt{cm}^{2}$ during the stress changing from 8 GPa to -8 GPa. The changing trend agrees with the results in Ref. [\cite{r31}], which investigated the change of polarization in $\pm3\%$ strained BiFeO$_{3}$, but is opposite to the results in Ref. [\cite{r17}], which studied the change of polarization under uniaxial stress in PbTiO$_{3}$. An earlier report \cite{r9} reveals that the polarization contribution arising from displacement of Fe and O atoms are almost canceling out and more than 98$\%$ of the net polarization present in BiFeO$_{3}$ is contributed by the Bi ions. Another report \cite{r29} indicates that the polarization comes from the lone electron pair of Bi$^{3+}$ ions. So, it is reasonable to deduce that the near-linearly variational polarization is related to (i) the displacement and (ii) the Born effective charge of Bi ions. And it is indeed interesting to analyse the relative displacement and Born effective charge of Bi ions in BiFeO$_{3}$ in detail under diferent uniaxial stress.

\begin{figure}
%\begin{figure}[tbp]
\begin{center}
\includegraphics[width=0.9\columnwidth]{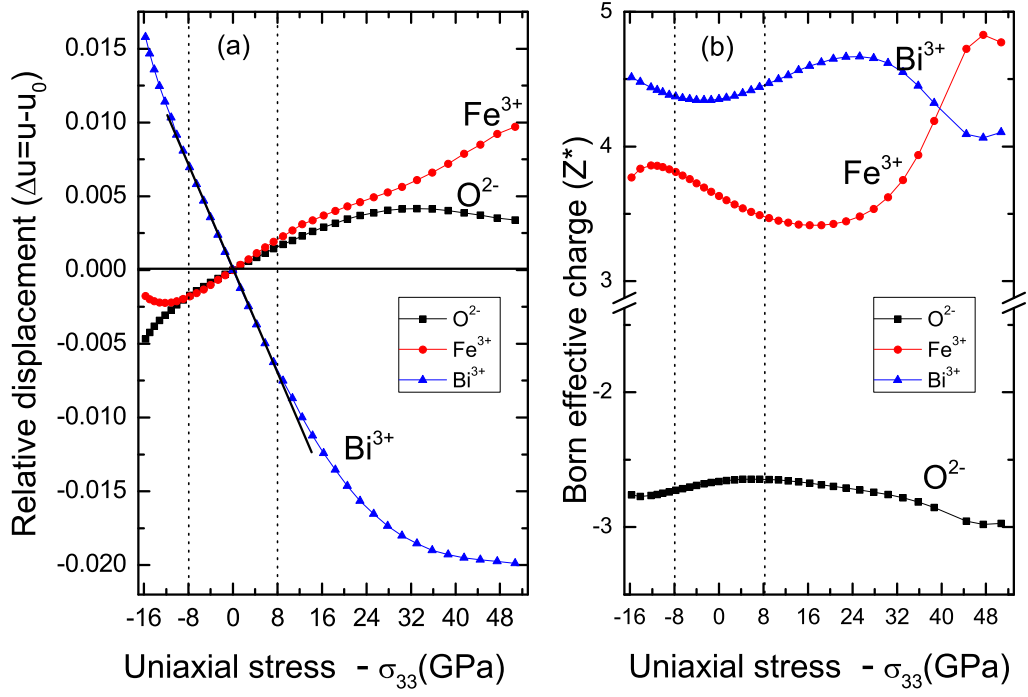}
\end{center}

\caption{\label{fig3}
(Color online) (a) Relative displacement ($\Delta$$u$ = $u - u_{0}$) and (b) Born effective charge of ions in the cell as a function of uniaxial stress -$\sigma_{33}$, where $u_{0}$ is the position of ions in unstrained BiFeO$_{3}$ under fractional coordinates.}
\end{figure}

Let us now concentrate on the relative displacement ($\Delta$$u$ = $u$ - $u$$_{0}$) of ions in the cell (under fractional coordinates) vs. uniaxial stress. Strikingly, Fig. \ref{fig3}(a) reveals that the changing curve of relative displacement of Bi$^{3+}$ ions is almost perfect linear in the uniaxial stress range of 8 GPa to -8 GPa and the moving direction of Bi$^{3+}$ ions happen to be [001] direction (polar axis, not shown), which are compatible with the near-linearly variational polarization. Note that the curves of Fe$^{3+}$ ion and O$^{2-}$ ion are almost overlap in the uniaxial stress range of 8 GPa to -8 GPa ,which indicates that the FeO$_{6}$ octahedron is not destroyed in this stress range. In addition, not only Bi$^{3+}$ ions, but also Fe$^{3+}$ and O$^{2-}$ ions have near-linearly changing curve of relative displacement in the uniaxial stress range of 8 GPa to -8 GPa. But the curve-slopes of Fe$^{3+}$ and O$^{2-}$ ions are far less than Bi$^{3+}$ ions', and the changing direction of relative displacement between Bi$^{3+}$ ions and Fe$^{3+}$ (or O$^{2-}$) ions in the cell is opposite (because of opposite slopes), which implys that the largest relative displacements are those of Bi$^{3+}$ ions relative to O$^{2-}$ (or Fe$^{3+}$) ions, as revealed in most reports of unstrained BiFeO$_{3}$ \cite{r3,r9,r29}.

To get more detailed information on near-linearly variational polarization, the Born effective charge (Z*) of ions in BiFeO$_{3}$ were obtained in the stress range of 16 GPa to -52 GPa , shown in Fig. \ref{fig3}(b). Interestingly, the $|$Z*$|$ (absolute value of Z*) of Fe$^{3+}$ and O$^{2-}$ ions decrease synchronously with the uniaxial stress changing from 8 GPa to -8 GPa, associating with the results of relative displacement of Fe$^{3+}$ and O$^{2-}$ ions demonstrated in Fig. \ref{fig3}(a), which show that the polarization contribution coming from Fe and O atoms are still almost canceling out and the net polarization present in BiFeO$_{3}$ is mainly contributed by the Bi ions. However, it should be noted that the Z* of Bi$^{3+}$ ions decrease first and then increase, not always increase, with the uniaxial stress changing from 8 GPa to -8 GPa. So, although the Bi$^{3+}$ ions have a biggish relative displacement in the range of 8 GPa to -8 GPa, the polarization of BiFeO$_{3}$ increases slowly. This result is in line with the earlier reports \cite{r31,r32} and give a possible explanation for it.

\subsection{Elastic properties under uniaxial stress}

\begin{figure}
%\begin{figure}[tbp]
\begin{center}
\includegraphics[width=0.9\columnwidth]{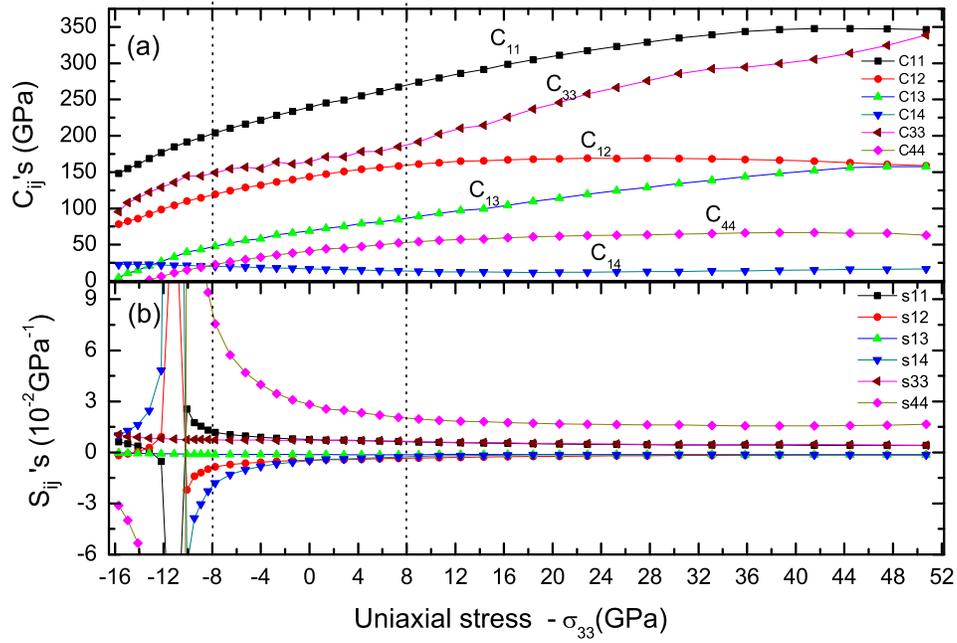}
\end{center}

\caption{\label{fig4}
(Color online) (a) Elastic stiffness coefficients $c_{ij}$'s and (b) elastic compliance coefficients $s_{ij}$'s as a function of uniaxial stress -$\sigma_{33}$.}
\end{figure}

The elastic stiffness coefficients ($c_{ij}$'s) and elastic compliance coefficients ($s_{ij}$'s) as a function of uniaxial stress (-$\sigma_{33}$) are shown in Figs. \ref{fig4}(a) and \ref{fig4}(b). The $c_{ij}$'s and $s_{ij}$'s collectively referred to as elasticity coefficient. They represent whether a material is easy to be stretched or not. The bigger the $c_{ij}$'s, the stronger the resisting deformation ability. On the contrary, the bigger the $s_{ij}$'s, the easier a material to be stretched. According to the symmetry of $R3c$ space group, the $c_{ij}$'s only have six independent variables\cite{r33}, namely, $c_{11}$, $c_{12}$ , $c_{13}$, $c_{14}$, $c_{33}$, $c_{44}$ (The $s_{ij}$'s have six similar variables, viz. $s_{11}$, $s_{12}$ , $s_{13}$, $s_{14}$, $s_{33}$, $s_{44}$). All of them can be obtained through the relationship between strain and stress. It must be noted that both $c_{ij}$'s and $s_{ij}$'s should satisfy the Born stability criterions. If not, the structure is unstable \cite{r33}. For a rhombohedral structure, the Born stability criterions are

\begin{equation}
c_{11} - |c_{12}| > 0,
\label{eq1}
\end{equation}
\begin{equation}
(c_{11} + c_{12})c_{33} - 2c_{13}^{2} > 0,
\label{eq2}
\end{equation}
\begin{equation}
(c_{11} - c_{12})c_{44} - 2c_{14}^{2} > 0.
\label{eq3}
\end{equation}
The form of constraint condition of $s_{ij}$'s is the same as the above formulas (not shown). Figure \ref{fig4}(a) shows that, except for $c_{14}$, the other $c_{ij}$'s increase near-linearly when the stress change from 8 GPa to -8 GPa. When the tensile stress is bigger than 11 GPa, there is a great change in $c_{ij}$'s: except for $c_{14}$, the other $c_{ij}$'s decrease rapidly and tend to 0. It is obvious that the structure is mechanically unstable at larger tensile stress based on the Born stability criterions Eq. (\ref{eq1})-(\ref{eq3}). For example, according to our calculation, the elastic stiffness coefficients are $c_{11}$ = 185 GPa, $c_{12}$ = 104 GPa, $c_{13}$ = 33 GPa, $c_{14}$ = 22 GPa, $c_{33}$ = 135 GPa and $c_{44}$ = 11 GPa when the tensile stress is 11.12 GPa. These coefficients can not meet the Eq. (\ref{eq3}). So the structure is predicted to be elastic instability, due to the decrease in $c_{44}$ and the increase in $c_{14}$. This predicted instability could suggest a ferroelastic phase transition \cite{r34} or melting \cite{r35} achieved at negative pressures (such as in epitaxial films) or high temperatures. These results are consistent with an earlier report of BiFeO$_{3}$ under hydrostatic pressure \cite{r7}. The difference is that all the $c_{ij}$'s of AFM BiFeO$_{3}$ are very sensitive to the hydrostatic pressure (see Fig . 2 of Ref. \cite{r7}). But when we focused on the $c_{ij}$'s in Fig. \ref{fig4}(a), we only found $c_{33}$ increases unstably and relative-rapidly due to the high uniaxial compressive stress in the [001] direction.

Similarly, Fig. \ref{fig4}(b) shows that, except for $s_{44}$, the absolute values of $s_{ij}$'s decrease near-linearly during the stress changing from 8 GPa to -8 GPa. When the tensile stress is bigger than 11 GPa, the absolute values of $s_{11}$, $s_{12}$, $s_{14}$ and $s_{44}$ increase rapidly. These great change make $s_{ij}$'s no longer satisfy the Born stability criterions \cite{r33} and imply that the structure is elastic instability, which is tally with the above analysis.

\subsection{Structural properties under uniaxial stress}

\begin{figure}
%\begin{figure}[tbp]
\begin{center}
\includegraphics[width=0.9\columnwidth]{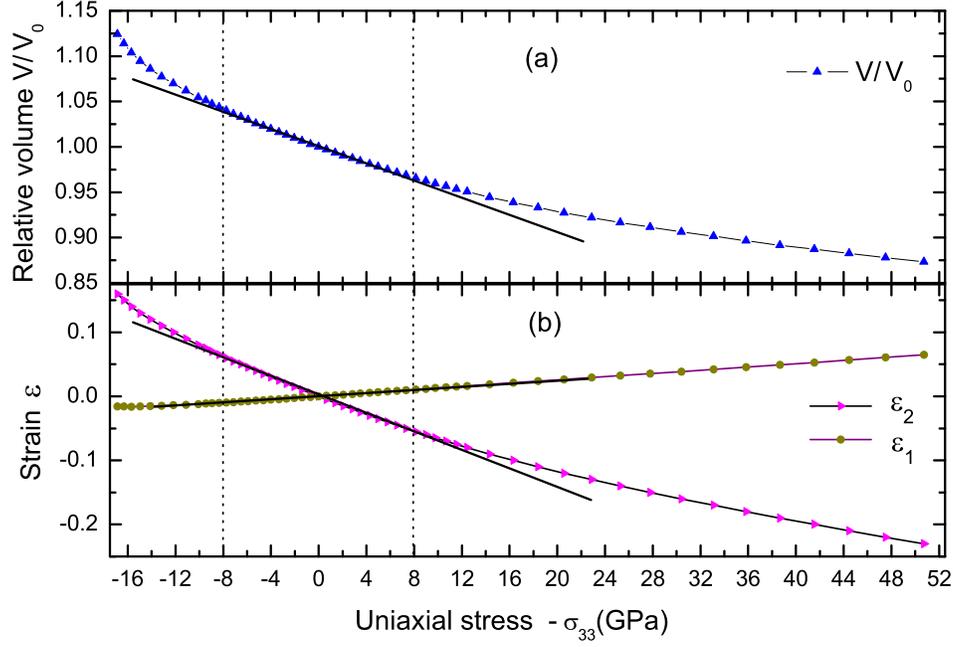}
\end{center}

\caption{\label{fig5}
(Color online) (a) $V/V_{0}$, (b) $\varepsilon_{1}$ = $(a/a_{0} - 1)$ and $\varepsilon_{2}$ = $(c/c_{0} - 1)$ as a function of uniaxial stress -$\sigma_{33}$, where $V_{0}$ and $a_{0}$, $c_{0}$ are the volume and lattice parameters of unstrained BiFeO$_{3}$.}
\end{figure}

From the analysis of elastic properties of BiFeO$_{3}$ under uniaxial stress, we found that the structure of BiFeO$_{3}$ is stable within the stress range of 8 GPa to -8 GPa, and when the tensile stress is greater than 11 GPa, the structure of BiFeO$_{3}$ becomes unstable obviously. This predicted structural stability from the analysis of elastic properties should be reflected in the change of structural parameters. To validate this idea, we analyzed the relative volume $V/V_{0}$ and the strains $\varepsilon_{1} = (a/a_{0}-1)$, $\varepsilon_{2} = (c/c_{0}-1)$ as a function of uniaxial stress, shown in Fig. \ref{fig5}. It is obvious that the $V/V_{0}$, $\varepsilon_{1}$ and $\varepsilon_{2}$ change near-linearly during the stress changing from 8 GPa to -8 GPa. Their slopes are -$4.64\times10^{-3}$, $1.67\times10^{-3}$ and -$7.16\times10^{-3}$, respectively. These phenomena indicate that the structure is stable in this uniaxial stress range. However, the relationship between $V/V_{0}$ (or $\varepsilon_{2}$) and uniaxial stress becomes nonlinear as the tensile stress increases, implying that the strain has crossed the elastic limit (or has a ferroelastic phase transition \cite{r34}) and the structure is mechanically unstable. The results are consistent with the above analysis of elasticity and provide direction for future applications.

\section{Conclusions}

We have studied the changes of polarization, elastic properties and structural stability of BiFeO$_{3}$ with rhombohedral \emph{G}-type AFM structure under uniaxial stress. First, we compare our results with available measurements and predictions. Then, we observe that the polarization \texttt{P} increases almost linearly from 75.45 $\mu \texttt{C}/\texttt{cm}^{2}$ to 97.08 $\mu \texttt{C}/\texttt{cm}^{2}$ during the stress changing from 8 GPa to -8 GPa and we have analyzed its two possible causes: (i) the displacement and (ii) the Born effective charge of Bi ions. Third, we study the elastic properties and structural stability comprehensively and find that the structure is stable in the stress range of 8 GPa to -8 GPa. These investigates provide helpful guidance for the applications of BiFeO$_{3}$, and will lead to more extensive studies, such as the changing rules of elastic properties, magnetic properties, polarization and phase transition under biaxial stress in BiFeO$_{3}$ thin films, and so on.

\section{Acknowledgements}

J. Li gratefully acknowledges financial support from the National Science Fund for Distinguished Young Scholar (Grant No. 60925016). This work is supported by the National Basic Research Program of China (Grant No. 2011CB921901) and the External Cooperation Program of Chinese Academy of Sciences.

%%%%%%%%%%%%%%%%%%%%%%%%%%%%%%%%%%%%%%%%%%%%%%%%%%%
\clearpage
%%%%%%%%%%%%%%%%%%%%%%%%%%%%%%%%%%%%%%%%%%%%%%%%%%%

\begin{table}
%\begin{table}[tbp]
\caption{\label{tab1}
The lattice parameters, polarization and the elastic stiffness coefficients of unstrained BiFeO$_{3}$: our calculated results, theoretical value in Ref. [\cite{r7,r9,r29}]and experimental values. The units of $a$ (or $c$), $\textbf{P}$, and $c_{ij}$'s are angstrom ({\AA}), $\mu \texttt{C}/\texttt{cm}^{2}$ and GPa, respectively.}

\begin{ruledtabular}
\begin{tabular}{lccccccccc}
         & $a$   & $c$    & $\textbf{P}$  & $c_{11}$  & $c_{12}$   & $c_{13}$  & $c_{14}$  & $c_{33}$  & $c_{44}$ \\
\hline
 \textbf{Our} & \textbf{5.56}  & \textbf{13.77}  & \textbf{87.5}  & \textbf{239}  & \textbf{144}  & \textbf{69}
              & \textbf{17}    & \textbf{164}    & \textbf{41} \\
 Theory & 5.51\footnotemark[1]	 & 13.54\footnotemark[1]	 & 88.7\footnotemark[2]/93.3\footnotemark[1]
     	& 222\footnotemark[3]	 & 110\footnotemark[3]	 & 50\footnotemark[3]	 & 16\footnotemark[3]	
        & 150\footnotemark[3]	 & 49\footnotemark[3] \\
 Experiment	& 5.57\footnotemark[4]	& 13.86\footnotemark[4]	& 59-90\footnotemark[5]/100\footnotemark[6] \\

\end{tabular}
\end{ruledtabular}
\footnotetext[1]{First-principles prediction by LSDA+$U$, $U_{eff}$ = 4 eV (Ref. \cite{r29}).}
\footnotetext[2]{First-principles prediction by GGA (Ref. \cite{r9}).}
\footnotetext[3]{First-principles prediction by GGA+$U$, Ueff = 6 eV (Ref. \cite{r7}).}
\footnotetext[4]{Reference \cite{r28}.}
\footnotetext[5]{Thin films (400 - 100 nm) on SrRuO$_{3}$/SrTiO$_{3}$ (Ref. \cite{r8}).}
\footnotetext[6]{Reference \cite{r30}.}

\end{table}

%%%%%%%%%%%%%%%%%%%%%%%%%%%%%%%%%%%%%%%%%%%%%%%%%%%

%%%%%%%%%%%%%%%%%%%%%%%%%%%%%%%%%%%%%%%%%%%%%%%%%%%

%%%%%%%%%%%%%%%%%%%%%%%%%%%%%%%%%%%%%%%%%%%%%%%%%%%

%%%%%%%%%%%%%%%%%%%%%%%%%%%%%%%%%%%%%%%%%%%%%%%%%%%

%%%%%%%%%%%%%%%%%%%%%%%%%%%%%%%%%%%%%%%%%%%%%%%%%%%

%%%%%%%%%%%%%%%%%%%%%%%%%%%%%%%%%%%%%%%%%%%%%%%%%%%

%%%%%%%%%%%%%%%%%%%%%%%%%%%%%%%%%%%%%%%%%%%%%%%%%%%

\end{document}